\documentclass[12pt]{article}
\usepackage[left=2.5cm,top=2.50cm,right=2.5cm,bottom=2.50cm]{geometry}
\usepackage{mathrsfs}
\usepackage{amsmath,amssymb,latexsym,color,cancel,graphicx,bbm,colortbl}
\usepackage[english]{babel}
\usepackage[latin1]{inputenc}
\usepackage{ragged2e}
\usepackage{cite}
\usepackage{subfig}
\begin{document}
\date{}

\title{Exact Solutions of the Schr\"odinger-Dunkl Equation for a Free Particle in a Finite and Infinite Cylindrical Well}

\author{ R. D. Mota$^{a}$, D. Ojeda-Guill\'en$^{b}$\footnote{{\it E-mail address:} dojedag@ipn.mx}\\ and
M. Salazar-Ram\'{\i}rez$^{b}$ }\maketitle

\begin{minipage}{0.9\textwidth}

\small $^{a}$ Escuela Superior de Ingenier{\'i}a Mec\'anica y El\'ectrica, Unidad Culhuac\'an,
Instituto Polit\'ecnico Nacional, Av. Santa Ana No. 1000, Col. San
Francisco Culhuac\'an, Alc. Coyoac\'an, C.P. 04430, Ciudad de M\'exico, Mexico.\\

\small $^{b}$ Escuela Superior de C\'omputo, Instituto Polit\'ecnico Nacional,
Av. Juan de Dios B\'atiz esq. Av. Miguel Oth\'on de Mendiz\'abal, Col. Lindavista,
Alc. Gustavo A. Madero, C.P. 07738, Ciudad de M\'exico, Mexico.\\

\end{minipage}

\begin{abstract}

In this paper, we study the Schr\"odinger equation with Dunkl derivative for a free particle confined in a cylindrical potential well. We consider both the finite and infinite height cases. The Dunkl formalism introduces reflection operators that modify the structure of the Hamiltonian and affect the parity of the solutions. By working in cylindrical coordinates, we obtain exact analytical expressions for the radial and axial wavefunctions in terms of Bessel functions. The energy spectrum and the solutions are classified according to the eigenvalues of the reflection operators in the three coordinates. We analyze in detail the conditions under which the wavefunctions acquire definite parity and discuss the resulting constraints on the Dunkl parameters.

\end{abstract}


\section{Introduction}

Over the last decades, differential operators involving reflection symmetries, known as Dunkl derivatives, have provided a fruitful generalization of standard quantum mechanical frameworks. These operators, which combine partial derivatives with reflections tied to finite reflection groups, were originally inspired by early ideas from Wigner \cite{wigner} and Yang \cite{yang}, and later formalized by Dunkl \cite{dunkl1,dunkl2}. The Dunkl derivative has been applied to solve several important problems in quantum mechanics, such as the harmonic oscillator and the Coulomb potential. In these models, the presence of reflection operators modifies the structure of the Hamiltonian and leads to changes in the energy spectrum and the symmetry of the wavefunctions. These types of systems have also been analyzed using algebraic methods, and they often exhibit superintegrability or hidden symmetries \cite{GEN1,GEN2,GEN3,GEN4,nos1,nos2,sami1,sami2}. The Dunkl derivative has also been applied in the relativistic context, where it has been used to solve both the Klein-Gordon and Dirac equations. In particular, several studies have analyzed systems such as the Coulomb potential, the Klein-Gordon oscillator, and the Dirac-Moshinsky oscillator within this framework \cite{nos3,nos4,nos5}.

The Dunkl derivative continues to be an active area of research in mathematical physics. In recent years, many works have explored its applications to exactly solvable models, quantum algebras, special functions, and relativistic equations. These studies have extended the use of Dunkl operators to higher-dimensional systems, new types of potentials, and generalized symmetries. Several research groups are currently working on this topic, and recent contributions can be found in Refs.
\cite{Quesne,Junker,Junker2,Schulze,Schulze2,Schulze3,Arab,Hassanabadi,Benzair,Raber,Benarous,Hamil, Bouguerne1,Bouguerne2,Benchikha,Lut, Hocine,Rou,Hamil2,Hamil3}.

On the other hand, the study of a free particle confined in a cylindrical potential well is a classical problem in quantum mechanics. Depending on the boundary conditions, the cylinder can have finite or infinite height, and both versions allow the analysis of quantized states in bounded regions. Recently, the infinite cylindrical well has attracted theoretical attention, as discussed in Refs. \cite{shihai,shihai2}. In contrast, the finite cylindrical well has been investigated both theoretically and experimentally, with applications in areas such as quantum dots, nanowires, and the electronic properties of low-dimensional systems \cite{baltenkov}. In particular, the cylindrical potential well has been used to describe anisotropic effects and confinement in heterostructures and optical systems. Moreover, there are two important limiting cases for this model. When the cylinder radius is very small compared to its height, it provides a one-dimensional model of the electrons along the cylinder. The other limiting case is when the cylinder height is very small compared to its radius, we obtain a model for the motion of electrons essentially on a disk \cite{baltenkov}.

Although the solution to the Schr\"odinger equation in this geometry is well known, its generalization using the Dunkl derivative is still not studied. Cylindrical wells, which are simple yet versatile in traditional quantum mechanics, offer a compelling setting to explore how reflection operators modify the structure of the energy spectrum and wavefunction behavior. In this paper, we study a free quantum particle confined within a cylindrical well of finite and infinite height by solving the Schr\"odinger equation generalized in terms of the Dunkl derivatives. We provide exact analytical solutions for both the radial and axial parts of the wavefunction, and carefully analyze how the parity imposed by the reflection operators constrains the allowed states. The classification of the solutions is carried out in terms of the eigenvalues of the reflection operators acting on each coordinate.

This paper is organized as follows. In Section 2, we introduce the generalized Schr\"odinger equation with the Dunkl derivative in cylindrical coordinates. In Section 3, we solve the equation for a free particle confined in a finite cylindrical potential well. We obtain the energy spectrum and the corresponding wavefunctions, and we carry out a detailed analysis of the parity of the states. In Section 4, we extend the results to the case of an infinite cylindrical well and present the corresponding exact solutions. Finally, we summarize our main findings and discuss possible future directions.

\section{The Schr\"odinger Equation Generalized with Dunkl Derivative }

The standard Dunkl derivative is defined as
\begin{equation}
 D_i\equiv\frac{\partial}{\partial x_i}+\frac{\mu_i}{x_i}(1-R_i), \hspace{5ex} i=1,2,3.\label{dunkl1}
\end{equation}
The constants $\mu_i$, which are called the Dunkl parameters, satisfy $\mu_i>-1/2$, $i=1, 2, 3$\cite{GEN4}, and $R_i$ are the reflection operators with respect to the $x-$, $y-$ and $z-$coordinates. Thus, $R_1f(x,y,z)=f(-x,y,z)$, $R_2f(x,y,z)=f(x,-y,z)$ and $R_3f(x,y,z)=f(x,y,-z)$.
With these  definitions ${\bf P}$ changes to ${\bf P}^2\equiv-\nabla^2_{D}$, where
\begin{equation}
\nabla^2_{D}=D_1^2+D_2^2+D_3^2  \label{DGEN}
\end{equation}
is the generalized Dunkl-Laplacian. Hence, if we set $\hbar=m=1$, the stationary generalized Schr\"odinger-Dunkl equation takes the form
\begin{equation}
H\Psi\equiv\left(-\frac{1}{2}{\nabla }_{D}^2+V(x,y,z)\right)\Psi=E\Psi.\label{DS}
\end{equation}

The action of the reflection operator $R_i$ on a three variables function $f(x,y,z)$ implies
\begin{eqnarray}
&&R_i^2=1, \hspace{3ex}\frac{\partial}{\partial x_i}R_i=-R_i\frac{\partial}{\partial x_i},\hspace{3ex}R_ix_i=-x_iR_i,\hspace{3ex}R_iD_i=-D_iR_i, \hspace{3ex} \hbox{$ i=x, y, z$}\label{pro1}\\
&&R_iR_j=R_jR_i, \hspace{3ex}\frac{\partial}{\partial x_i}R_j=R_j\frac{\partial}{\partial x_i},\hspace{3ex}R_ix_j=x_jR_i,\hspace{3ex}R_iD_j=D_jR_i, \hspace{3ex} \hbox{$i\ne j$}.
\end{eqnarray}
Also, the following equalities  can be proved
\begin{eqnarray}
&&R_iR_j=R_jR_i,\hspace{5ex}[ D_i,  D_j]=0,\\
&&[x_i,{D}_i]=-(1+2\mu_i) R_{i}.\label{pro2}
\end{eqnarray}

Proceeding as in Ref. \cite{GEN3}, it is immediate to write the Schr\"odinger equation (\ref{DGEN}), for a free particle $V(x,y,z)=0$, in  polar coordinates as
\begin{equation}
\left(A_\rho+\frac{1}{\rho^2}B_\phi+C_z \right) \Psi(\rho,\phi,z)=E  \Psi(\rho,\phi,z), \label{laplapol}
\end{equation}
where the operators $A_\rho$, $C_z$, and  $B_\phi$ are given by
\begin{eqnarray}
&&A_\rho=-\frac{1}{2}\left(\frac{\partial^2}{\partial \rho^2}+\frac{1+2\mu_1+2\mu_2}{\rho}\frac{\partial }{\partial \rho}\right),\\
&&C_z=-\frac{1}{2}\left(\frac{\partial ^2}{\partial z^2}+\frac{2\mu_3}{z}\frac{\partial }{\partial z}-\frac{\mu_3}{z^2}(1-R_3)\right),\\
&&B_\phi\equiv-\frac{1}{2}\frac{\partial^2}{\partial \phi^2}+\left(\mu_1\tan{\phi}-\mu_2\cot{\phi}\right)\frac{\partial}{\partial \phi}
+\frac{\mu_1 (1-R_1)}{2\cos^2{\phi}}+\frac{\mu_2 (1-R_2)}{2\sin^2{\phi}}.\label{bfi}
\end{eqnarray}
To obtain the result of equation (\ref{laplapol}) the following properties were used
\begin{equation}
R_1f(\rho,\phi,z)=f(\rho,\pi-\phi,z),\hspace{2ex}R_2f(\rho,\phi,z)=f(\rho,-\phi,z),\hspace{2ex}R_3f(\rho,\phi,z)=f(\rho,\phi,-z).
\end{equation}
For the purpose of the present paper, we propose the separation of variables of the wavefunction as
\begin{equation}
\Psi(\rho,\phi,z)=R(\rho)\Phi(\phi)\psi(z),
\end{equation}
with the eigenvalue equations
\begin{eqnarray}
&&C_z\psi(z)=\epsilon_z\psi(z),\label{eqnz} \\
&&B_\phi\Phi(\phi)=\frac{s^2}{2}\Phi(\phi). \label{angulareigen}
\end{eqnarray}
This implies that the radial equation takes the form
\begin{equation}
\left(A_\rho+\frac{s^2}{2 \rho^2}+\epsilon_z\right)R(\rho)=\epsilon R(\rho).\label{antes}
\end{equation}
The angular equation have been solved, and its  eigenfunctions $\Phi_\ell^{(e_1,e_2)}(\phi)$ and eigenvalues  $\frac{s^2}{2}$ have been found by Genest {\it et. al.} \cite{GEN3}. Here, ($e_1,e_2$) correspond to the eigenvalues ($1-2e_1,1-2e_2)$$\equiv(r_1,r_2)$ of the reflection operators ($R_1,R_2$), see Appendix.
It is worth highlighting that, for all parity cases, the angular separation constant $\frac{s^2}{2}$ takes the same form \cite{GEN3}
\begin{equation}
s^2=4\ell(\ell+\mu_1+\mu_2), \label{scuad}
\end{equation}
with $\ell = 0, 1, 2, 3, ...$ for $ (r_1,r_2)=(1,1)$ or  $ (r_1,r_2)=(-1,-1)$. Similarly, $\ell = \frac{1}{2}, \frac{3}{2}, \frac{5}{2}, ... $ for $ (r_1,r_2)=(1,-1)$ or  $ (r_1,r_2)= (-1,1)$. All these results are reported in Table 1 for clarity. A very important fact that should be emphasized is that the product of the eigenvalues of the reflection operators $r_1\times r_2$ is +1 for $\ell=0, 1, 2, 3, ...$ and -1 for $\ell=\frac{1}{2}, \frac{3}{2}, \frac{5}{2}, \frac{7}{2}, ...$

\begin{table}[ht]
\begin{center}
\begin{tabular}{| c | c|c | c |}\hline \hline
$(e_1,e_2) $& $(r_1,r_2)$& $r_1\times r_2$& $\ell$\\ \hline \hline
(0,0), (1,1) & (1,1), (-1,-1)&1 & 0, 1, 2, 3, ...\\ \hline
(0,1), (1,0) & (1,-1), (-1,1)&-1& $\frac{1}{2}, \frac{3}{2}, \frac{5}{2}, \frac{7}{2}, ...$\\ \hline \hline
\end{tabular}
\caption{It shows the values $(e_1,e_2)$ corresponding to the eigenvalues $(1-2e_1,1-2e_2)\equiv (r_1,r_2)$ of the parity operators $(R_1,R_2)$ in the $x-y$ plane. Also reports the product of the eigenvalues $r_1\times r_2$ as well as their corresponding angular eigenvalues.}
\label{l table}
\end{center}
\end{table}
The result presented in (\ref{scuad}) allows us to write equation (\ref{antes}) as follows
\begin{equation}
\left(\frac{d^2}{d \rho^2}+\frac{1+2\mu_1+2\mu_2}{\rho}\frac{d}{d \rho}-\frac{4\ell(\ell+\mu_1+\mu_2)}{\rho^2}+2(\epsilon-\epsilon_z)\right)R(\rho)=0.\label{RadialSD}
\end{equation}
In the remainder of this work, we focus on solving equations (\ref{eqnz}) and (\ref{RadialSD}) and analyzing the parity properties of the wavefunctions.

\section{The Finite Dunkl Cylindrical Well Free-Particle Solutions}

\subsection{The $z$-coordinate solutions}

We are now in a position to solve the equations (\ref{eqnz}) and (\ref{RadialSD}) for a free particle in the cylindrical potential well. This means that the potential is zero inside the cylinder and infinite outside it. Specifically, we choose a cylinder with radius $R_c$ and height $2H$, centered on the origin of the coordinates, with its axis of symmetry on the $z$-axis (see Fig. 1). The chosen position of the cylinder ensures that coordinate reflections are well defined and is based on the choice that has been made for the infinite one-dimensional potential well studied with Dunkl derivative \cite{chungx,chung2}. Since the particle is confined inside the cylinder, the wavefunctions $\psi(z)$ and $R(\rho)$ must obey the boundary conditions $\psi(\pm H)=0$ and $R(R_c)=0$, which we will impose below to obtain the energy spectrum.\\

\begin{figure}[ht]
 \centering
    \includegraphics[width=0.45\textwidth]{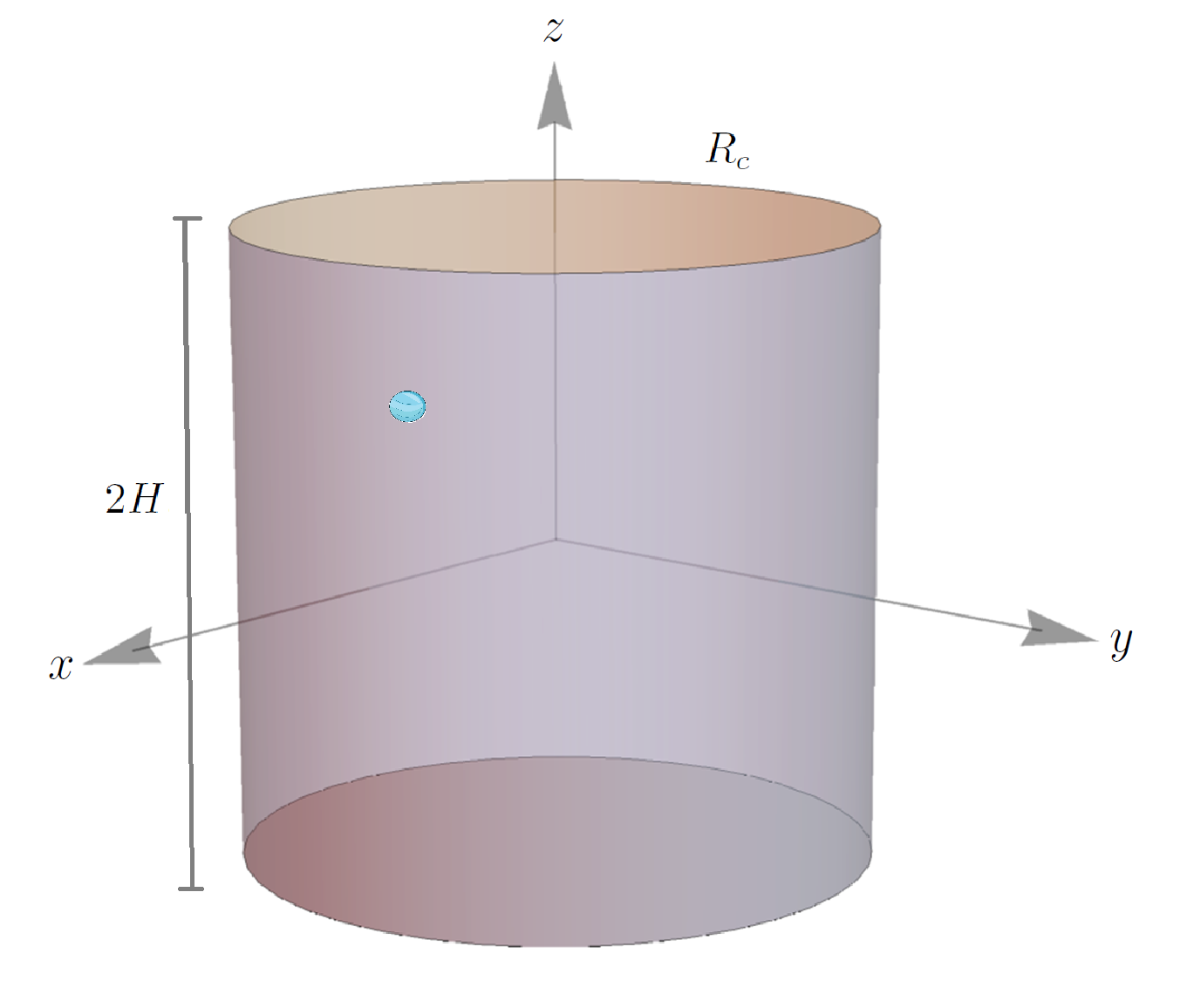}
    \caption{\footnotesize Schematic representation of a vertical cylinder of radius $R_c$ and height $2H$, aligned along the $z$-axis. The origin of the coordinate system is located at the center of the cylinder. This geometry is used to model the spatial confinement of a free particle in cylindrical coordinates.}
\end{figure}

{\it Case I: Even solutions}\\

By imposing the condition  $R_3\psi^+(z) =+\psi^+(z)$, equation (\ref{eqnz}) reduces to
\begin{equation}
\left(-\frac{1}{2}\frac {d ^2}{d z^2}-\frac{\mu_3}{z} \frac {d}{d z}\right)\psi^+(z) =\epsilon^+_z \psi^+(z).
\end{equation}
Since $\epsilon^+_z >0$  we introduce the change of variable $u\equiv\sqrt{2\epsilon^+_z}z$. Thus, this equation transforms to
\begin{equation}
\left(\frac {d ^2}{d u^2}+\frac{2\mu_3}{u} \frac {d}{d u} + 1\right)\psi^+(u)=0, \label{bessel1}
\end{equation}
which is a Lommel's type equation.

The Lommel's equation
\begin{equation}
v''(\xi)+\frac{1-2\alpha}{\xi}v'(\xi)+\left(\left( \beta\gamma \xi^{\gamma-1}\right)^2+\frac{\alpha^2-\nu^2\gamma^2}{\xi^2}\right)v(\xi)=0,\label{lommel}
\end{equation}
has the regular solutions at the origin
\begin{equation}
v(\xi)=\xi^\alpha J_\nu(\beta \xi^\gamma),
\end{equation}
being $ J_\nu(\xi)$ the first kind Bessel functions of order $\nu$  \cite{lebedev,nikiforov}.

Thus, by direct comparison of equations  (\ref{bessel1}) and (\ref{lommel}), we find
\begin{eqnarray}
&&\gamma=1,\hspace{8ex}\alpha=\frac{1}{2}-\mu_3,\\
&&\beta=1,\hspace{8ex}\nu=\mu_3-\frac{1}{2}.
\end{eqnarray}
Therefore, $ \psi^+(z)$ is given by
\begin{equation}
\psi^+(z)=C_e z^{-\mu_3+\frac{1}{2}}J_{\mu_3-\frac{1}{2}}\left(\sqrt{2\epsilon^+_z}z\right).\label{tipo1}
\end{equation}
The quantization of the energy spectrum follows from applying the boundary condition $\psi^+(\pm H)=0$. Let $\omega_{\mu_3-\frac{1}{2},n'}$ denote the $n'$th zero of the Bessel function $J_{\mu_3-\frac{1}{2}}(x)$, $n'=1, 2, 3,...$. Then, the allowed energy levels are
\begin{equation}
\epsilon_{z\,n'}^+=\frac{\omega_{\mu_3-\frac{1}{2}\,, n' }^2}{2H^2}.\label{espectromas}
\end{equation}
Values of the Bessel function zeros can be found in classical references \cite{jahnke,abramowitz}, as well as in more recent works \cite{shihai,baltenkov}. However, they can be computed efficiently using symbolic software. For instance, in Maple, the command {\it BesselJZeros(-0.3,3)}, it returns the third zero of $J_{-0.3}(x)$, which is  8.177851519. An important fact is that the Bessel functions $J_\nu(x)$ have an infinite number of zeros $x_{\nu,n}$, satisfying  $J_\nu(x_{\nu,n})=0$, $n=1, 2, 3,...$ \cite{jackson}.\\

{\it Case II: Odd solutions}\\

By imposing the condition $R_3\psi^-(z) =-\psi^-(z)$ into equation (\ref{eqnz}), we obtain
\begin{equation}
\left(-\frac{1}{2}\frac {d ^2}{d z^2} -\frac{\mu_3}{z} \frac {d}{d z}+\frac{\mu_3}{z^2}\right)\psi^-(z) =\epsilon^-_z \psi^-(z).
\end{equation}
With the change of variable  $v\equiv\sqrt{2\epsilon^-_z}z$, this equation can also be brought to the Lommel's type equation
\begin{equation}
\left(\frac {d ^2}{d v^2}+\frac{2\mu_3}{v} \frac {d}{d v} -\frac{2\mu_3}{v^2}+ 1\right)\psi^-(v)=0. \label{bessel2}
\end{equation}
The comparison of this equation with equation (\ref{lommel}), leads us to find
\begin{eqnarray}
&&\gamma=1,\hspace{8ex}\alpha=\frac{1}{2}-\mu_3,\\
&&\beta=1,\hspace{8ex}\nu=\mu_3+\frac{1}{2}.
\end{eqnarray}
Hence the regular solutions  at the origin $ \psi^-(z)$ are given by
\begin{equation}
 \psi^-(z)=C_o z^{-\mu_3+\frac{1}{2}}J_{\mu_3+\frac{1}{2}}\left(\sqrt{2\epsilon^-_z}z\right).\label{tipo2}
\end{equation}
Also, in this case the boundary condition $\psi^-(\pm H)=0$ implies that the allowed energy levels are determined by the zeros of the Bessel function $J_{\mu_3+\frac{1}{2}}(x)$, denoted by $\omega_{\mu_3+\frac{1}{2}\,, n' }$, such that
\begin{equation}
\epsilon_{z\, n'}^-=\frac{\omega_{\mu_3+\frac{1}{2}\,, n' }^2}{2H^2}. \label{espectromenos}
\end{equation}
These results complete the determination of the axial energy spectrum and wavefunctions for both parity cases.

\subsection{Radial solutions}

We now turn to the radial equation (\ref{RadialSD}). By introducing the change of variable $w=\sqrt{2(\epsilon-\epsilon_z)}\rho$, it is straightforward to show that it can be written as
\begin{equation}
\left(\frac {d ^2}{d w^2}+\frac{1+2\mu_1+2\mu_2}{w} \frac {d}{d w} -\frac{2\ell(\ell+\mu_1+\mu_2)}{w^2}+ 1\right)R(w)=0. \label{besselradial}
\end{equation}
By comparison of this equation with the Lommel's equation (\ref{lommel}), we obtain
\begin{eqnarray}
&&\gamma=1,\hspace{8ex}\alpha=-(\mu_1+\mu_2),\\
&&\beta=1,\hspace{8ex}\nu=2\ell+\mu_1+\mu_2.
\end{eqnarray}
Therefore, we find that the regular radial solutions at the origin are given by
\begin{equation}
R(\rho)=C_\rho{\rho}^{-(\mu_1+\mu_2)}J_{2\ell+\mu_1+\mu_2}\left(\sqrt{2(\epsilon-\epsilon_z)}\rho\right).\label{fullradial}
\end{equation}
Since the particle is confined within a cylinder of radius $R_c$, the radial solutions must vanish on the surface of the cylinder. In analogy with the axial case, let $\Omega_{2\ell+\mu_1+\mu_2\,, n }$ the $n$th zero of the Bessel function $J_{2\ell+\mu_1+\mu_2}(x)$, $n=1, 2, 3, ...$. This condition leads to the quantization of the radial energy levels, and the total energy spectrum is given by
\begin{equation}
\epsilon_{n n'}=\frac{\Omega^2_{2\ell+\mu_1+\mu_2, n }}{2R_c^2}+\epsilon_{zn'}.\label{full}
\end{equation}
A complete description of the energy spectrum and the corresponding wavefunctions requires a more detailed analysis of the parity conditions, which will be addressed in the next subsection.

\subsection{Parity analysis}

It is important to note that, although we imposed the parity conditions $R_3\psi^\pm(z) =\pm \psi^\pm(z)$ on the solutions given in equations (\ref{tipo1}) and (\ref{tipo2}), these functions are not strictly even or odd in the  conventional sense. As it is shown in previous works involving the Dunkl derivative \cite{SCH,chung2,physicaA}, the wavefunctions do not possess strict parity under the action of the reflection operator unless specific conditions are satisfied, especially concerning the Dunkl parameters. In order for our solutions to strictly satisfy the parity requirements, we recall that only Bessel functions of integer order have well-defined parity. Specifically, they obey the following properties
\begin{equation}
R_3(J_p(z))=J_p(-z)=(-1)^pJ_p(z), \hspace{6ex}R_3(z^qJ_p(z))=(-1)^{q+p}J_p(z   ), \label{prop}
\end{equation}
where $q, p=0, \pm 1, \pm 2, \pm 3,...$. This shows that the parity of the function $x^qJ_p(x)$ is determined by the sum $q+p$. Additionally, it is known that for $q=-p, -p+1, -p+2,...$, that is, $p+q=0, 1, 2,...$, these functions are regular at the origin. From now on we will denote the eigenvalues of the reflection operator $R_3$ as $r_3$.

Thus, for the solutions (\ref{tipo1}) to have the desired even parity, the quantity $\mu_3-\frac{1}{2}$ must be an integer. If we define this integer as $m\equiv \mu_3-\frac{1}{2}$, since $\mu_3>\frac{1}{2}$, this implies $\mu_3-\frac{1}{2}>-1$. Then, $\mu_3=\frac{1}{2}, \frac{3}{2}, \frac{5}{2}, ...$. This selection automatically implies that the states  (\ref{tipo1}) are of the even form $z^{-m}J_m(z)$, with  $m=\mu_3-\frac{1}{2}=0,1, 2, ...$ A similar analysis leads us to obtain the condition  $\mu_3=\frac{1}{2}, \frac{3}{2}, \frac{5}{2}, ...$ for the odd wave solutions of equation  (\ref{tipo2}). This set of $\mu_3$ leads the odd solutions  (\ref{tipo2})  to have the form $z^{-m}J_{m+1}(z)$,  $m=\mu_3-\frac{1}{2}=0,1, 2, ...$

Consequently, we have that the correct energy spectra and wavefunctions in the $z$-coordinate, for $r_3=1$ (even parity) are given by
\begin{eqnarray}
&&\epsilon_{z m n'}^+=\frac{\omega_{m, n' }^2}{2H^2},\hspace{5ex} m=0,1 , 2, ...\label{espectropar}\\
&&\psi^{+}_{m n'}(z)=C_e z^{-m}J_{m}\left(\frac{\omega_{m, n' }}{H} z\right),
\end{eqnarray}
while for odd parity $r_3=-1$, they are given by
\begin{eqnarray}
&&\epsilon_{z m  n'}^-=\frac{\omega_{m+1, n' }^2}{2H^2}, \hspace{5ex} m=0,1 , 2, ...\label{espectroimpar}\\
&&\psi^-_{m n'}(z)=C_o z^{-m}J_{m+1}\left(\frac{\omega_{m+1, n' }}{H}z\right).\label{zimpar}
\end{eqnarray}

On the other hand, although the radial wavefunctions do not explicitly depend on parity, it is essential to ensure that the total wavefunction has a well-defined parity in each coordinate direction. Our analysis follows the parity structure summarized in Table 1.

Let us first consider the cases $(r_1,r_2)=(1,1)$ and $(-1,-1)$, for which the allowed values of $\ell$ are $\ell = 0, 1, 2, \dots$. In these configurations, the product of the reflection eigenvalues in the $x$- and $y$-coordinates is even, so the radial part of the wavefunction,
\begin{equation}
R(\rho) = \rho^{-(\mu_1 + \mu_2)} J_{2\ell + \mu_1 + \mu_2}(\sqrt{2(\epsilon - \epsilon_z)}\, \rho),
\end{equation}
must also be an even function. Let us define the Bessel index as $N \equiv 2\ell + \mu_1 + \mu_2$, and the Dunkl parameter sum as $M \equiv \mu_1 + \mu_2$. If $N$ is even, i.e., $N = 0, 2, 4, \dots$, then the function $\rho^\delta J_N(\rho)$ is even provided that $\delta + N$ is even. Since $\delta = -M$, we require $-M + N \equiv N - M$ to be even. This implies that $M$ must be even: $M = 0, 2, 4, \dots$. On the other hand, if we assume $N$ is odd $(N = 1, 3, 5, \dots )$, then the same argument leads to $M = 1, 3, 5, \dots$.

Now consider the cases $(r_1, r_2) = (1,-1)$ and $(-1,1)$, which correspond to \( \ell = \frac{1}{2}, \frac{3}{2}, \frac{5}{2}, \dots \). In these configurations, the product of the reflection eigenvalues is odd, and thus the radial wavefunctions must be an odd function. Using the same form of the solution and definitions of $N$ and $M$ as before, we apply the parity condition:
\begin{itemize}
    \item If $N = 0, 2, 4, \dots$, then to ensure odd parity we must have $M = 1, 3, 5, \dots$.
    \item If $N = 1, 3, 5, \dots$, then we require $M = 0, 2, 4, \dots$.
\end{itemize}
According to the introduced notation, $-M+N=2\ell=0, 2, 4,...$ or $-M+N=2\ell=1, 3, 5,...$, which is in full agreement with what is said below the equation (\ref{prop}). Both conditions can be written as $N-M\ge 0$. Also, the radial wavefunctions take the form
\begin{equation}
R_{NMn}(\rho) = \rho^{-M} J_{N}(\sqrt{2(\epsilon - \epsilon_z)}\, \rho).
\end{equation}

The above classification ensures that the full wavefunction respects the parity structure imposed by the reflection operators in each coordinate, and leads to consistent constraints on the Dunkl parameters $\mu_1$ and $\mu_2$ for all cases. All the previous results are summarized in Table 2. In this Table we show the allowed values of $N$ and $M$ and the eight distinct types of admissible states, each corresponding to a specific combination of eigenvalues $(r_1, r_2, r_3)$.

\begin{table}[ht]
\begin{center}
\begin{tabular}{|  p{22mm} | p{52mm}| c | c|}\hline \hline
$(r_1,r_2,r_3)$ & $N$, $\ell$, $M$, $m $ \hspace{2ex}$(N\ge M)$& Full coordinate wavefunctions \\ \hline \hline
(1,1,1) \hspace{34mm}(-1,-1,1) & $N$=0, 2, 4,...\hspace{4ex}$\ell= 0, 1, 2,...$\newline $M$=0, 2, 4,...\hspace{3ex}$m$=0, 1, 2,...& ${\rho}^{-M}J_{N}\left(\sqrt{2(\epsilon^+-\epsilon_z^+)}\rho\right)$ $\times$ $z^{-m}J_m\left(\sqrt{2\epsilon_z^+} z\right)$\\ \hline
(1,1,-1) \hspace{34mm}(-1,-1,-1) & $N$=1, 3, 5,...\hspace{4ex}$\ell= 0, 1, 2,...$\newline $M$=1, 3, 5,...\hspace{3ex}$m$=0, 1, 2,...& ${\rho}^{-M}J_{N}\left(\sqrt{2(\epsilon^+-\epsilon_z^-)}\rho\right)$ $\times$ $ z^{-m}J_{m+1}\left(\sqrt{2\epsilon_z^-} z\right)$\\ \hline
(1,-1,1) \hspace{34mm}(-1,1,1) & $N$=0, 2, 4,...\hspace{3ex}$\ell= \frac{1}{2},  \frac{3}{2} ,  \frac{5}{2}, ...$\newline $M$=1, 3, 5,...\hspace{3ex}$m$=0, 1, 2,...& ${\rho}^{-M}J_{N}\left(\sqrt{2(\epsilon^--\epsilon_z^+)}\rho\right)$ $\times$ $ z^{-m}J_m\left(\sqrt{2\epsilon_z^+} z\right)$\\ \hline
(1,-1,-1) \hspace{34mm}(-1,1,-1) & $N$=1, 3, 5,...\hspace{3ex}$\ell= \frac{1}{2},  \frac{3}{2} ,  \frac{5}{2},...$\newline $M$=0, 2, 4,...\hspace{3ex}$m$=0, 1, 2,...& ${\rho}^{-M}J_{N}\left(\sqrt{2(\epsilon^--\epsilon_z^-)}\rho\right)$ $\times$ $ z^{-m}J_{m+1}\left(\sqrt{2\epsilon_z^-} z\right)$\\ \hline\hline
\end{tabular}
\caption{Classification of the eigenvalues allowed of the reflection operators $R_1$, $R_2$, and $R_3$, along with the corresponding allowed values of the quantum number $\ell$, the sum of the Dunkl parameters in the $x$-$y$ plane $(M = \mu_1 + \mu_2)$, the third Dunkl parameter $(m = \mu_3 - \tfrac{1}{2})$, and the explicit form of the full wavefunctions for each case.}
\label{2_table}
\end{center}
\end{table}

\subsection{Full energy spectrum and wavefunctions}

We now apply the results obtained for the Dunkl solutions in the $z$-coordinate, where the energy eigenvalues depend on the parity of the $z$-wavefunctions. Since the radial solutions must have definite parity, equation (37) for an energy $\epsilon_z^+$ (or for an energy $\epsilon_z^-$) implies that there are two energies $\epsilon^\pm$, depending on the parity of the radial wavefunctions. Thus, the complete energy spectrum of a free particle within a finite potential well can be obtained by combining equations (\ref{full}), (\ref{espectropar}), and (\ref{espectroimpar}). Using the notation introduced in the previous section we can write the total energy as
\begin{equation}
  \epsilon^\pm_{N n,m n'}\equiv \epsilon_{\rho Nn}+\epsilon^\pm_{zmn'}=\left\lbrace
  \begin{array}{l}
      \frac{\Omega^2_{N,n }}{2R_c^2}+\epsilon^+_{zmn'}=\frac{\Omega^2_{N,n }}{2R_c^2}+\frac{\omega_{m,n' }^2}{2H^2},\\
      \frac{\Omega^2_{N,n }}{2R_c^2}+\epsilon^-_{zmn'}=\frac{\Omega^2_{N,n }}{2R_c^2}+\frac{\omega_{m+1,n' }^2}{2H^2},\label{totalspectrum}\\
  \end{array}
  \right.
\end{equation}
where we have substituted $N=2\ell+\mu_1+\mu_2$ into equation (\ref{full}), and $m=\mu_3-\frac{1}{2}$ into equations  (\ref{espectromas}) and (\ref{espectromenos}). Note that although we have generically defined the radial energy spectrum in equation (\ref{totalspectrum}) as
\begin{equation}
\epsilon_{\rho Nn}=\frac{\Omega^2_{N,n }}{2R_c^2},
\end{equation}
it must be emphasize that its values depend on the quantum numbers of the radial wavefunction in Table 2, that is, on whether the radial state is even or odd.
 From equation (\ref{totalspectrum}), we obtain that the multiplying factor of the variable $\rho$ of the argument of the radial wavefunctions in Table 2  is given by
\begin{equation}
 \sqrt{2\left(\epsilon^\pm_{N n,m n'}-\epsilon^\pm_{zmn'}\right)}=\frac{\Omega_{N,n }}{R_c}.
\end{equation}

As an illustrative example, consider the case $(r_1,r_2,r_3)=(1,-1,-1)$. That is, both wave functions, the radial and the $z$-coordinate, are odd. From the parity conditions in Table 2, this corresponds to $N=1, 3, 5, ...$, $N \ge M$, and the axial parity requires $m=0, 1, 2, ...$. Therefore, the energy spectrum $\epsilon^-_{N n,m n'}$, and the wavefunction denoted by $\Theta^{-}_{NM n, m n'}(\rho, z)$ take the form
\begin{eqnarray}
&&\epsilon^-_{N n,m n'}=\frac{\Omega^2_{N,n }}{2R_c^2}+\epsilon^-_{zmn'}=\frac{\Omega^2_{N,n }}{2R_c^2}+\frac{\omega_{m+1,n' }^2}{2H^2},\\
&&\Theta^{-}_{NMn,m n'}(\rho,z)=C^-{\rho}^{-M}J_{N}\left(\frac{\Omega_{N,n }}{R_c}\rho\right)z^{-m}J_{m+1}\left(\frac{\omega_{m+1, n' }}{H} z\right).
\end{eqnarray}
Hence, we have fully characterized the energy spectrum and the corresponding eigenfunctions under all parity configurations imposed by the reflection operators.

In Figure 2, we plot particular cases of the energy spectrum  $\epsilon_{\rho Nn}$ and $\epsilon^-_{zmn'}$. In Fig (a) we plot the radial energy $\epsilon_{\rho Nn}$ by setting  $R_c =10$, for the different $N=1,3,5,$ considered as a function of $n=1,2,3,4,5$. In Fig (b) we plot the $z$ energy $\epsilon^-_{zmn'}$ by setting  $H=15$, for the different $m=1,3,5,$ considered as a function of $n'=1,2,3,4,5$. Although these plots are for particular cases, they represent the typical behavior of the energies  $\epsilon_{\rho Nn}$ and $\epsilon^-_{zmn'}$, since with other quantum numbers they behave similarly.

\begin{figure}[ht]
 \centering
  \subfloat[]{
    \includegraphics[width=0.5\textwidth]{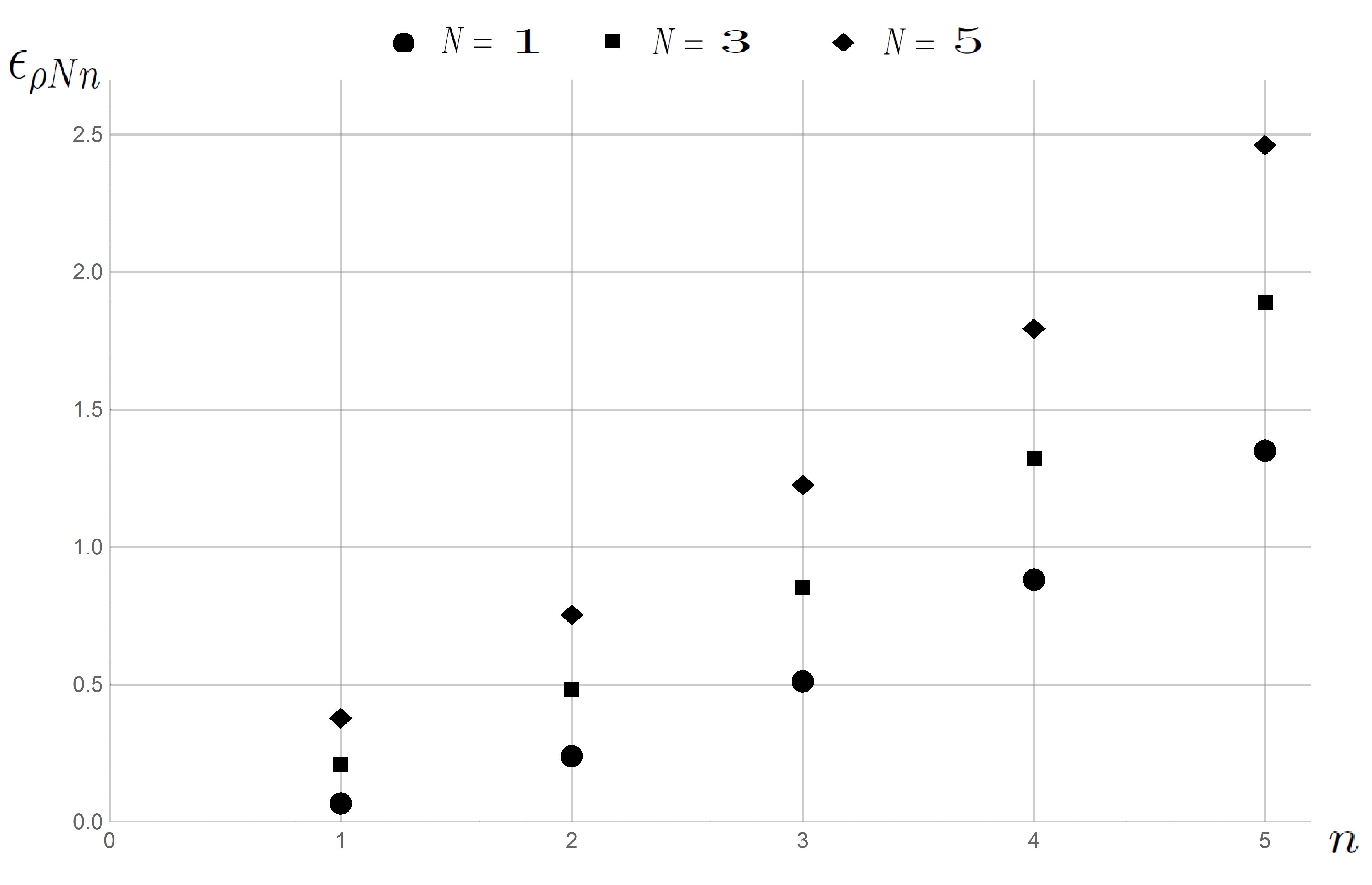}}
  \subfloat[]{
    \includegraphics[width=0.5\textwidth]{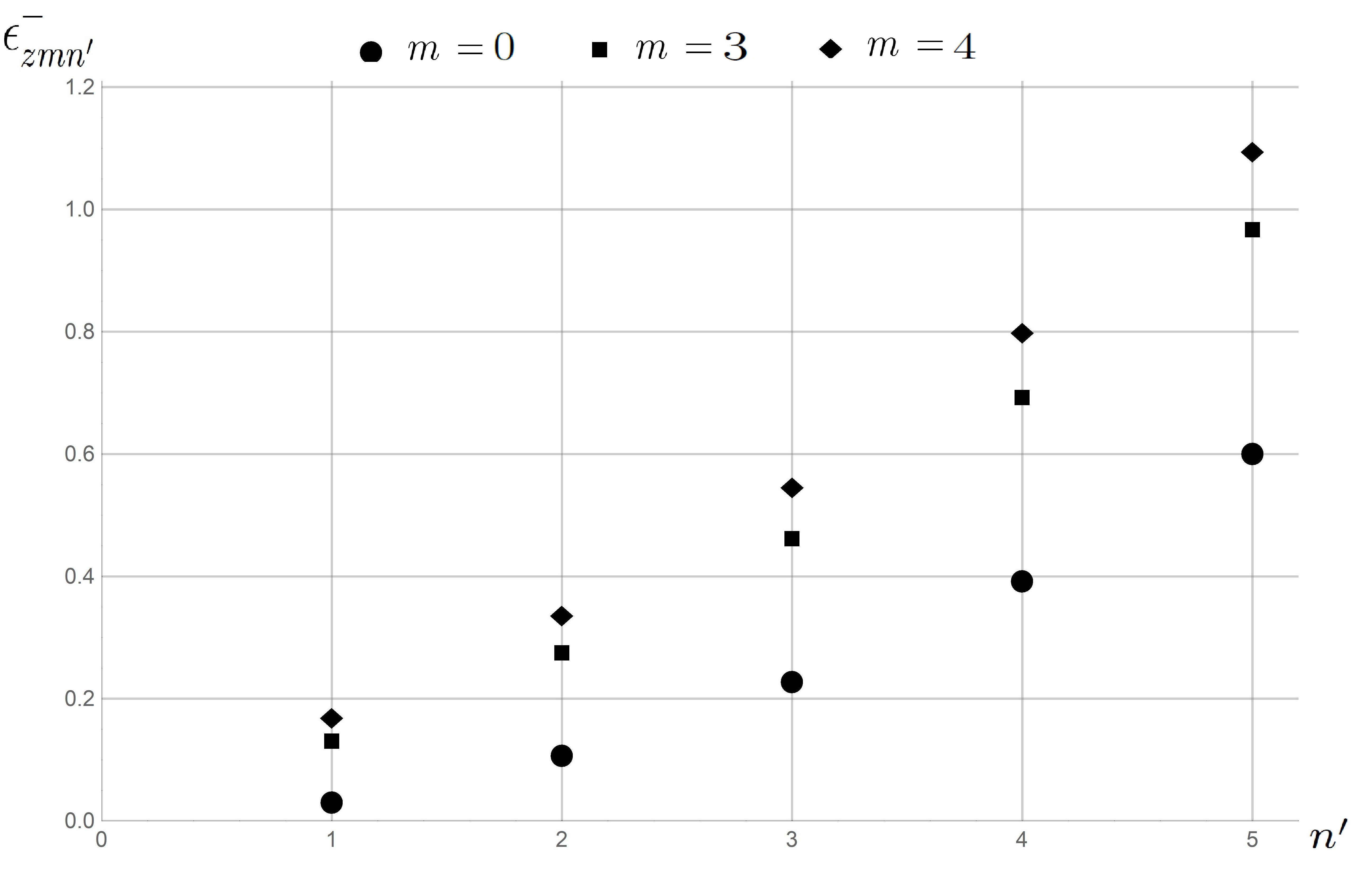}}
  \caption{\footnotesize Radial $\epsilon_{\rho Nn}$ and axial $\epsilon^-_{zmn'}$ energy spectra for selected quantum numbers. (a) Radial energies $\epsilon_{\rho Nn}$ for $R_c =10$ and $n=1,2,3,4,5$. The circles represent the energies with $N=1$, the squares represent the energies for $N = 3$ and the diamonds represents the energies for $N=5$. (b) Axial energies $\epsilon^-_{zmn'}$ for $H=15$ and $n'=1,2,3,4,5$. The circles represent the energies with $m=1$, the squares represent the energies for $m=3$ and the diamonds represents the energies for $m=5$.}
\end{figure}

In Figures 3 and 4, we plot unnormalized typical even and odd wavefunctions. In Figure 3, we plot some wavefunctions $R_{NMn}(\rho)={\rho}^{-M}J_{N}\left(\frac{\Omega_{N,n }}{R_c}\rho\right)$ for $R_C=10$. In Figure 4, we plot some even wavefunctions $\psi^+_{mn'}(z)={z}^{-m}J_m\left(\frac{\omega_{m,n'}}{H}\rho\right)$ and some odd wavefunctions  $\psi^-_{mn'}(z)={z}^{-m}J_m\left(\frac{\omega_{m+1,n'}}{H}\rho\right)$ for $H=15$.

\begin{figure}[ht]
 \centering
  \subfloat[]{
    \includegraphics[width=0.5\textwidth]{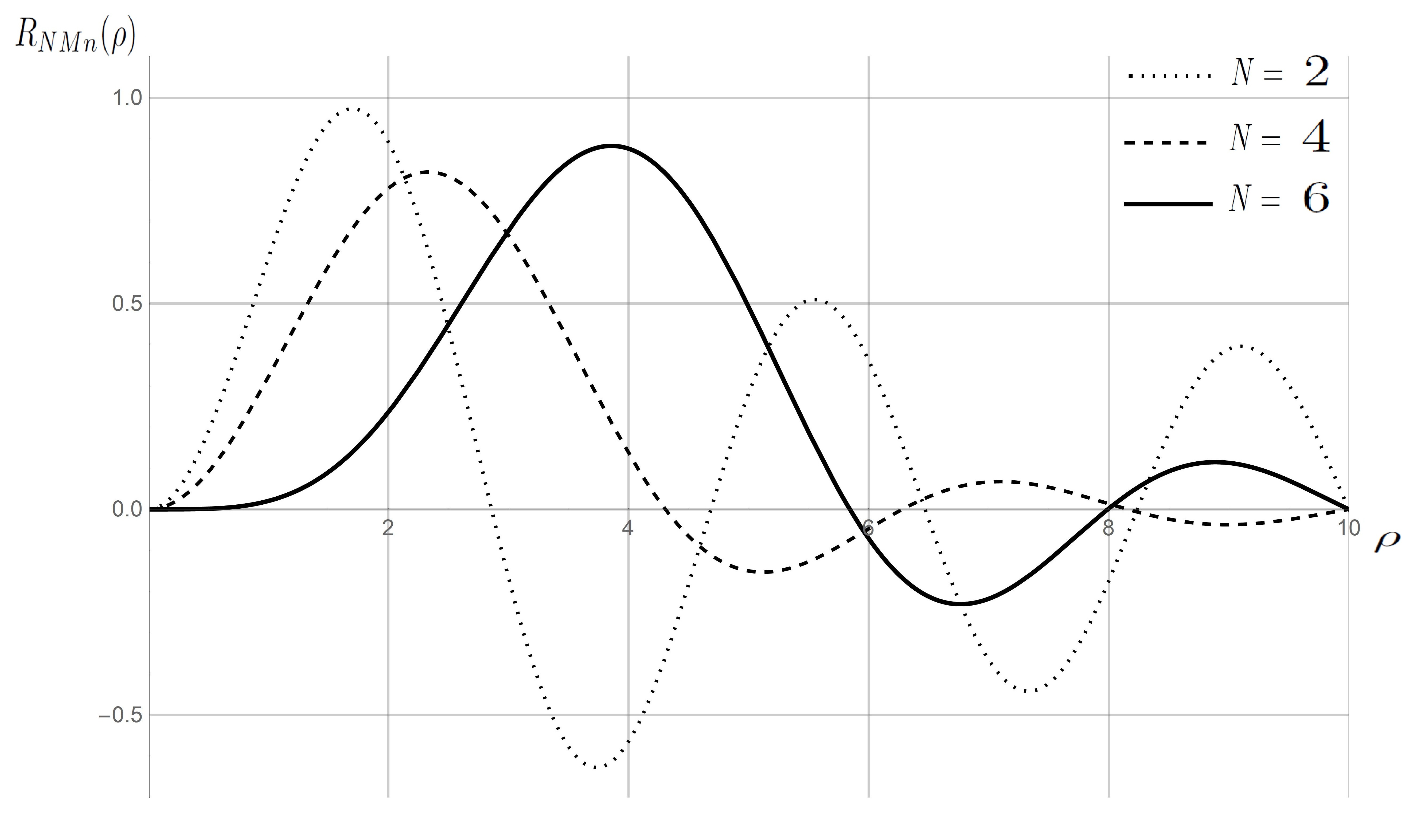}}
  \subfloat[]{
    \includegraphics[width=0.5\textwidth]{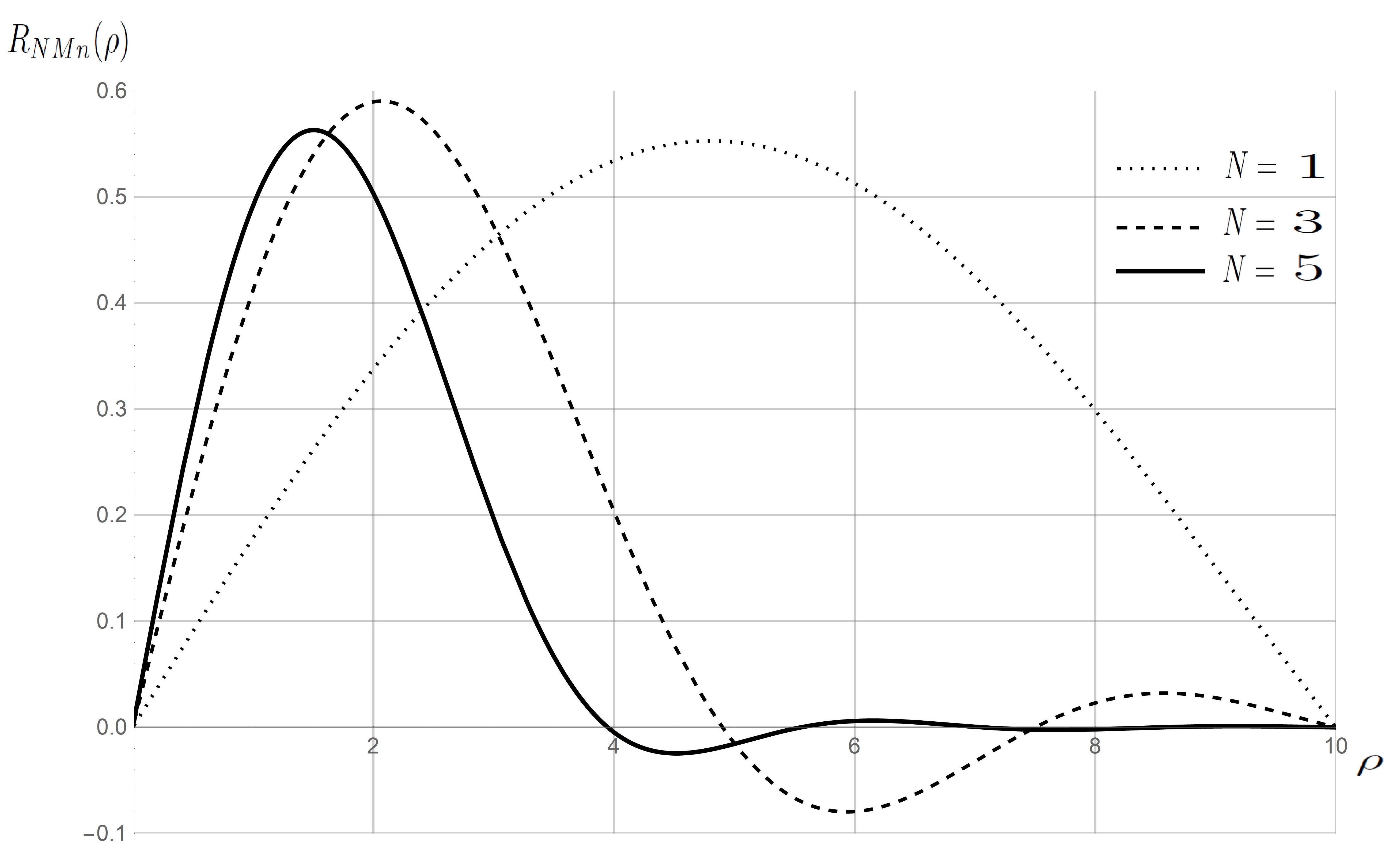}}
  \caption{\footnotesize Plots for some unnormalized radial wavefunctions $R_{NMn}(\rho)={\rho}^{-M}J_{N}\left(\frac{\Omega_{N,n }}{R_c}\rho\right)={\rho}^{-M}J_{N}\left(\frac{\Omega_{N,n }}{R_c}\rho\right)$ for $R_C=10$. (a) Even-parity radial wavefunctions for $N=2$, $M=0$, $\Omega_{2,5}$ (dot curve), for $N = 4$, $M=2$, $\Omega_{4,4}$ (dash curve) and $N=6$, $M=2$, $\Omega_{6,3}$ (solid curve). (b) Odd-parity radial wavefunctions for $N=1$, $M=0$, $\Omega_{1,1}$ (dot curve), for $N = 3$, $M=2$, $\Omega_{3,3}$ (dash curve) and $N=5$, $M=4$, $\Omega_{5,5}$ (solid curve). }
\end{figure}

\begin{figure}[ht]
 \centering
  \subfloat[]{
    \includegraphics[width=0.5\textwidth]{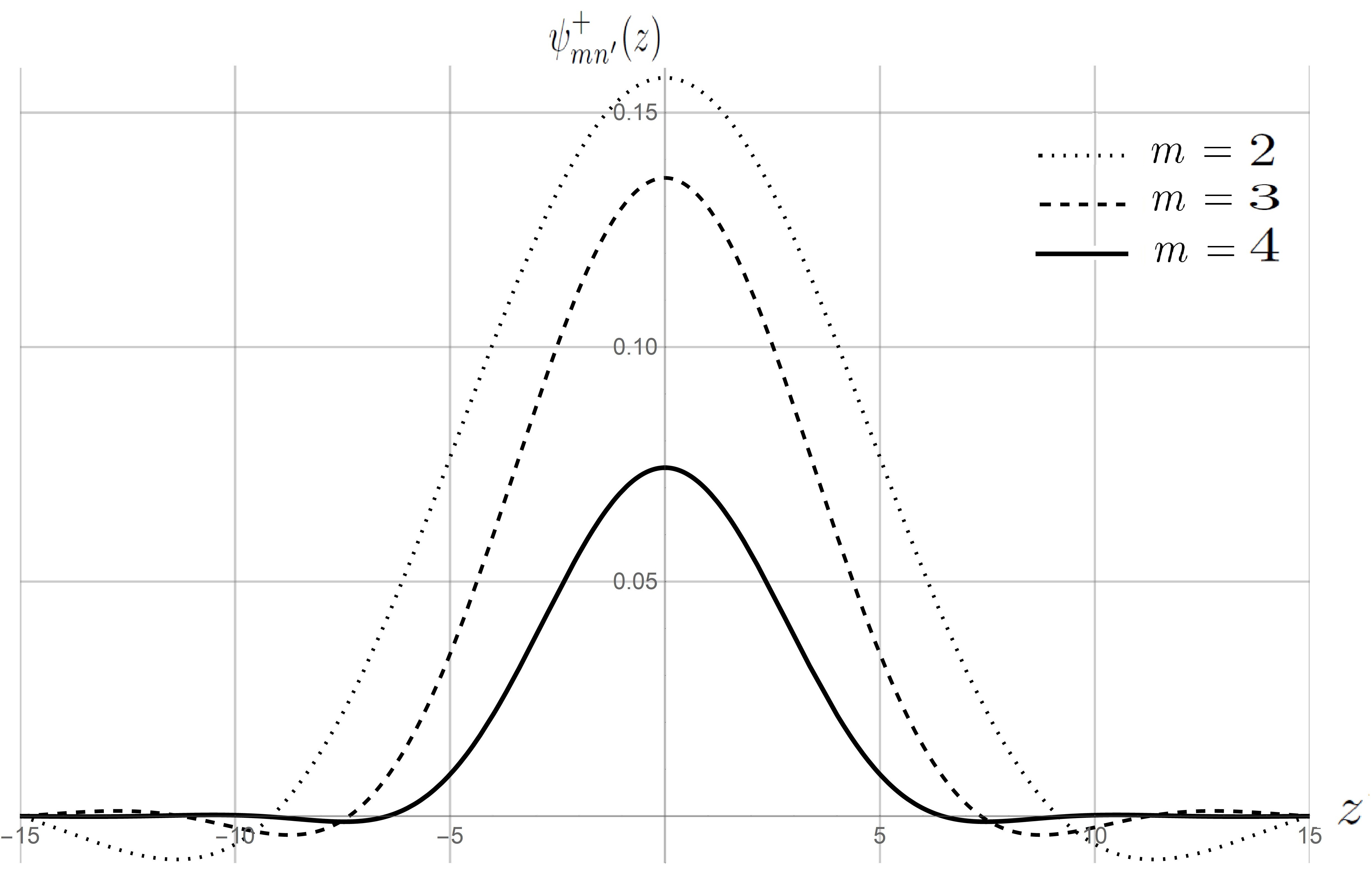}}
  \subfloat[]{
    \includegraphics[width=0.5\textwidth]{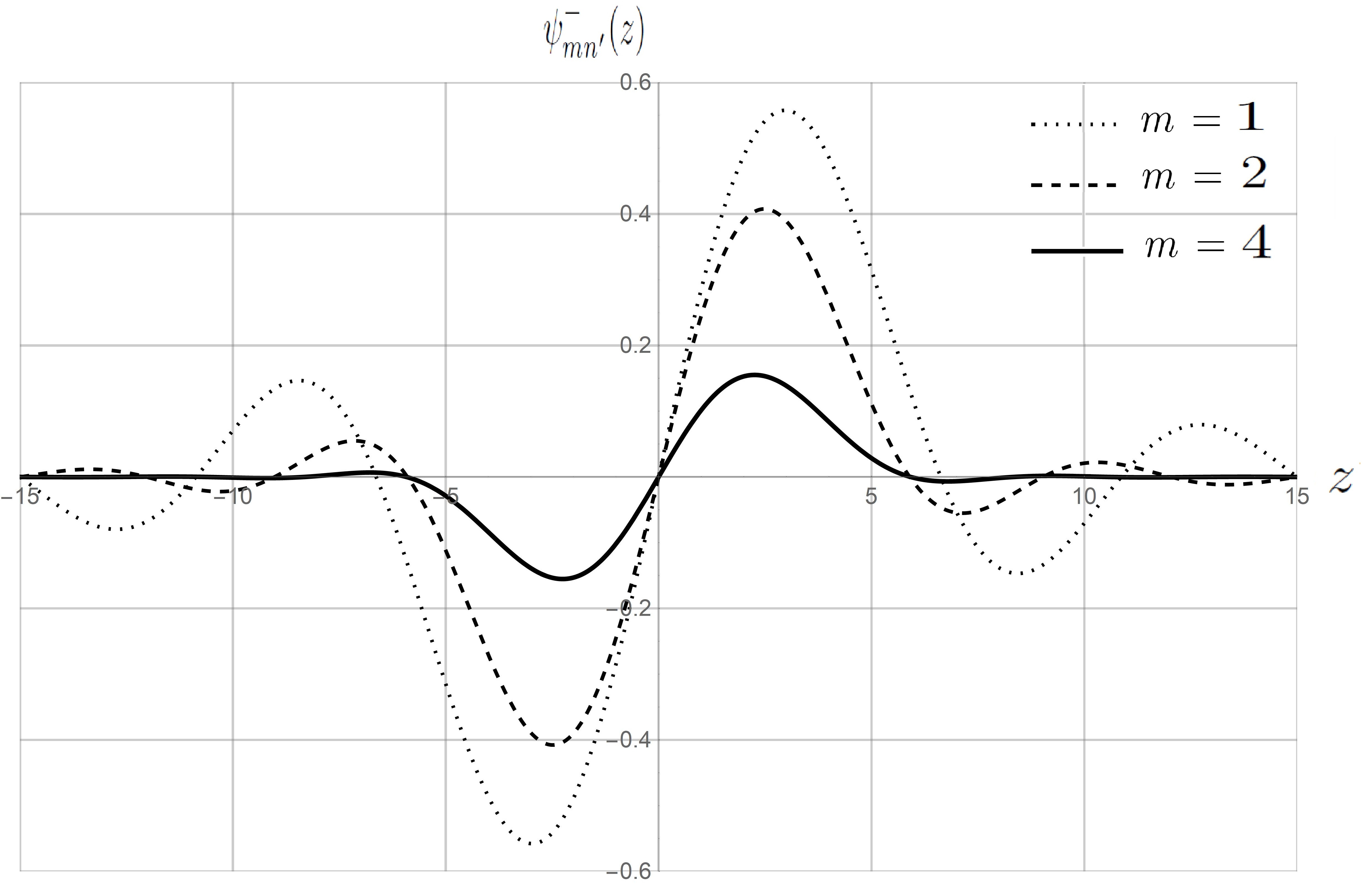}}
  \caption{\footnotesize Plots for some unnormalized even wavefunctions $\psi^+_{mn'}(z)={z}^{-m}J_m\left(\frac{\omega_{m,n'}}{H}z\right)$ and for some odd wavefunctions  $\psi^-_{mn'}(z)={z}^{-m}J_{m+1}\left(\frac{\omega_{m+1,n'}}{H}z\right)$ with $H=15$. In Fig (a) we plot the even wavefunctions for $m=2$, $\omega_{2,2}$ (dot curve), for $m= 3$, $\omega_{3,3}$ (dash curve) and $m=4$, $\omega_{4,4}$ (solid curve). In Fig (b) we plot the odd wavefunctions for $m=1$, $\omega_{2,3}$ (dot curve), for $m=2$, $\omega_{3,4}$ (dash curve) and $m=4$, $\omega_{5,5}$ (solid curve).}
\end{figure}

\section{The Infinite Dunkl Cylindrical Well Free-Particle Solutions}

In this case, the structure of the solutions in the $z$-coordinate, as well as the parity conditions, remain valid. Since the particle can move freely along the $z$-axis, then $\epsilon^\pm$ is the kinetic energy of the particle in that direction: $\epsilon^\pm =\frac{k^2}{2}$, with $k\in \mathbb{R}$. Thus, based on the results shown in equations (\ref{espectropar})-(\ref{zimpar}), we now obtain
\begin{eqnarray}
&&\epsilon^\pm_z=\frac{k^2}{2},\label{z1}\\
&&\psi^+(z)=C_e z^{-m}J_{m}\left(k z\right),\label{z2}\hspace{5ex}m=0, 1, 2, ...\\
&&\psi^-(z)=C_o z^{-m}J_{m+1}\left(k z\right), \label{z3}\hspace{3ex}m=0, 1, 2, ...
\end{eqnarray}
On the other hand, the radial solutions in the $x-y$ plane, together with the previously established parity conditions, remains unchanged. Consequently, the eight allowed wavefunctions listed in Table 2 are still applicable in this case. However, the solutions in the $z$-coordinate must be replaced by those given in equations (\ref{z2}) and  (\ref{z3}).

Now, from equation (\ref{totalspectrum}) we obtain that the total energy spectrum of the particle in the infinite cylinder well is given by
\begin{equation}
  \epsilon^\pm_{N n}=\frac{\Omega^2_{N,n }}{2R_c^2}+\epsilon^\pm=\frac{\Omega^2_{N, n}}{2R_c^2}+\frac{k^2}{2}.\hspace{1.2cm}
\end{equation}
This completes the analysis of the infinite cylindrical well. In the following section, we summarize our main results and outline possible directions for future research.

\section{Concluding Remarks}

In this work, we have solved the generalized Schr\"odinger equation with Dunkl derivatives for a free particle confined in a cylindrical potential well, considering both the finite and infinite height cases. To achieve this, we expressed the Dunkl-Laplacian in cylindrical coordinates and obtained exact analytical solutions for the radial and axial parts of the wavefunction. We found that the radial and axial components of the wavefunction can be written in terms of Bessel functions. We also established the necessary conditions on the Dunkl parameters to ensure that the solutions have well-defined parity. This study extends previous results related to non-isotropic confinement models and contributes to the understanding of quantum systems with discrete symmetries.

As a continuation of this research, we plan to explore similar systems using generalized forms of the Dunkl derivative \cite{chung1}. Another possible direction is to study the Schr\"odinger equation with Dunkl operators in spherical confinement, which corresponds to an isotropic case.

\section*{Acknowledgments}

This work was partially supported by SNII-M\'exico, COFAA-IPN, EDI-IPN, and CGPI-IPN Project Number 20251355.

\section*{Appendix}

It can be shown that the operator  $B_\phi$ defined in equation (\ref{bfi}) and the angular momentum operator around the $z$-axis, $L_z\equiv -i(xD_2-yD_1)$, are not independent, but are related through the expression
\begin{equation}
L_z^2=2B_\phi+2\mu_1\mu_2(1-R_1R_2).
\end{equation}
For a Hamiltonian of the form $H_C\equiv-\frac{1}{2}\left(D_1^2+D_2^2\right)+V(\rho)$, it can be shown that the reflection operators commute with the Hamiltonian, $ [R_1,H_C]=[R_2,H_C]=0$. In Refs. \cite{GEN1,GEN4}, the eigenfunctions $\Phi(\phi)$ and eigenvalues $\frac{s^2}{2}$ of the operator $B_\phi$ were determined, satisfying
\begin{equation}
B_\phi\Phi(\phi)=\frac{s^2}{2}\Phi(\phi)\label{esf}.
\end{equation}
The eigenfunctions $\Phi(\phi)$ are explicitly given by
\begin{equation}
\Phi_\ell^{(e_1,e_2)}(\phi)=\eta_\ell^{(e_1,e_2)}\cos^{e_1}\phi\sin^{e_2}\phi \hspace{0.2cm} P_{\ell-e_1/2-e_2/2}^{\mu_1-1/2+e_1,\mu_2-1/2+e_2}(-\cos{2\phi}),\label{angular}
\end{equation}
which are written in terms of the quantum numbers ($e_1,e_2$), corresponding to the eigenvalues of the reflection operators ($R_1,R_2$), ($1-2e_1,1-2e_2$), and the Jacobi polynomials $P_\ell^{(\alpha,\beta)} (x)$.

It was shown in Refs. \cite{GEN1,GEN4} that
\begin{equation}\label{mcuad}
s^2=4\ell(\ell+\mu_1+\mu_2),
\end{equation}
where the allowed values of $\ell$ are restricted according to
\begin{equation*}
  (e_1,e_2)\in\left\lbrace
  \begin{array}{l}
      \left\{(0,0),(1,1)\right\},\hspace{1.0cm}\text{$\ell=0, 1, 2,... ,$ }\\
      \left\{(1,0),(0,1)\right\},\hspace{1.0cm}\text{$\ell=\frac{1}{2}, \frac{3}{2}, \frac{5}{2},... $}.\\
  \end{array}
  \right.
\end{equation*}
On the other hand, $\eta_\ell^{(e_1,e_2)}$ is the normalization constant, and it is given by
\begin{align}
\eta_\ell^{(e_1,e_2)}=&\sqrt{\left(\frac{2\ell+\mu_1+\mu_2}{2}\right)\left(\ell-\frac{e_1+e_2}{2}\right)}\nonumber\times\\
&\sqrt{\frac{\Gamma\left(\ell+\mu_1+\mu_2+\frac{e_1+e_2}{2}\right)}{\Gamma\left(\ell+\mu_1+\frac{1+e_1-e_2}{2}\right)\Gamma\left(\ell+\mu_2+\frac{1+e_2-e_1}{2}\right)}}.
\end{align}
 Furthermore, using the orthogonality relations of the Jacobi polynomials, it was shown in Refs. \cite{GEN1,GEN4} that the angular wavefunctions $\Phi_\ell^{(e_1,e_2)}(\phi)$ satisfy
\begin{equation}
\int_0^{2\pi}\Phi_\ell^{(e_1,e_2)}(\phi)\Phi_{\ell'}^{(e'_1,e'_2)}(\phi)|\cos{\phi}|^{2\mu_1}|\sin{\phi}|^{2\mu_2}d\phi=\delta_{\ell,\ell'}\delta_{e_1,e'_1}\delta_{e_2,e'_2}.
\end{equation}
 This confirms the orthonormality of the eigenfunctions under the appropriate weight function.


\begin{thebibliography} {99}

\bibitem{wigner}E. Wigner, {\it Phys. Rev.} \textbf{77} (1950) 711.

\bibitem{yang}L.M. Yang, {\it Phys. Rev.} \textbf{84} (1951) 788.

\bibitem{dunkl1} C.F. Dunkl, {\it Trans. Am. Math. Soc.} \textbf{311} (1989) 167.

\bibitem{dunkl2}C.F. Dunkl, and Y. Xu, {\it Orthogonal polynomials of several variables, Encyclopedia of Mathematics and Its Applications, Vol. 81}, Cambridge University Press, Cambridge, 2001.

\bibitem{GEN1}V.X. Genest, M.E.H. Ismail, L. Vinet, and A. Zhedanov, {\it J. Phys. A.} \textbf{46} (2013) 145201.

\bibitem{GEN2}V.X. Genest, M.E.H. Ismail, L. Vinet, and A. Zhedanov, {\it Commun. Math. Phys.} {\bf329} (2014) 999.

\bibitem{GEN3}V.X. Genest, L. Vinet, and A. Zhedanov, {\it J. Phys. Conf. Ser.} \textbf{512} (2014) 012010.

\bibitem{GEN4}V.X. Genest, A. Lapointe, and L. Vinet, {\it Phys. Lett. A} \textbf{379} (2015) 923.

\bibitem{nos1}M. Salazar-Ram\'{\i}rez, D. Ojeda-Guill\'en, R.D. Mota, and V.D. Granados, {\it Eur. Phys. J. Plus} \textbf{132} (2017) 39.
kov
\bibitem{nos2}M. Salazar-Ram\'{\i}rez, D. Ojeda-Guill\'en, R.D. Mota, and V.D. Granados, {\it Mod. Phys. Lett. A} \textbf{33} (2018) 1850112.

\bibitem{sami1}S. Ghazouani, I. Sboui, M.A. Amdouni, and M.B. El Hadj Rhouma, {\it J. Phys. A: Math. Theor.} \textbf{52} (2019) 225202.

\bibitem{sami2}S. Ghazouani, and I. Sboui, {\it J. Phys. A: Math. Theor.} \textbf{53} (2019) 035202.

\bibitem{nos3}R.D. Mota, D. Ojeda-Guill\'en, M. Salazar-Ram\'irez, and V.D. Granados, {\it Ann. Phys.} \textbf{411} (2019) 167964.

\bibitem{nos4}R.D. Mota, D. Ojeda-Guill\'en, M. Salazar-Ram\'irez, and V.D. Granados, {\it  Mod. Phys. Lett. A}, \textbf{36} (2021) 2150171.

\bibitem{nos5}D. Ojeda-Guill\'en, R.D. Mota, M. Salazar-Ram\'irez, and V.D. Granados, {\it Mod. Phys. Lett. A} \textbf{35} (2020) 2050255.

\bibitem{Quesne} C. Quesne, {\it J. Phys. A: Math. Theor.} \textbf{56} (2023) 265203.

\bibitem{Junker} G. Junker, S.H. Dong, P. Sedaghatnia, W.S. Chung, and H. Hassanabadi, {\it Ann. Phys.} \textbf{454} (2023) 169336.

\bibitem{Junker2} G. Junker, {\it J. Phys. A: Math. Theor.} \textbf{57} (2024) 075201.

\bibitem{Schulze} A. Schulze-Halberg, {\it Phys. Scr.} \textbf{99} (2024) 075212.

\bibitem{Schulze2} A. Schulze-Halberg, and P. Roy, {\it J. Phys. A: Math. Theor.} \textbf{57} (2024) 225204.

\bibitem{Schulze3} A. Schulze-Halberg, {\it Mod. Phys. Lett. A} \textbf{39} (2024) 2450178.

\bibitem{Arab} A. Arabsaghari, H. Hassanabadi, and W.S. Chung, {\it Mod. Phys. Lett. A} \textbf{39} (2024) 2450117.

\bibitem{Hassanabadi} S. Hassanabadi, J. K\v{r}\'i\v{z}, B.C. L\"utf\"uo\u{g}lu, W.S. Chung, P. Sedaghatnia, and H. Hassanabadi, {\it Int. J. Theor. Phys.} \textbf{63} (2024) 323.

\bibitem{Benzair} H. Benzair, T. Boudjedaa, and M. Merad, {\it Phys. Scr.} \textbf{99} (2024) 055261.

\bibitem{Raber} D.E.M. Raber, H. Benzair, T. Boudjedaa, and M. Merad, {\it Phys. Scr.} \textbf{100} (2025) 015277.

\bibitem{Benarous} M. Benarous, A. Hocine, B.C. L\"utf\"uo\u{g}lu, and B. Hamil, {\it J. Stat. Mech.} \textbf{2025} (2025) 053102.

\bibitem{Hamil} B. Hamil, B.C. L\"utf\"uo\u{g}lu, and M. Merad, {\it  Phys. Scr.} \textbf{100} (2025) 035301.

\bibitem{Bouguerne1} H. Bouguerne, B. Hamil, B.C. L\"utf\"uo\u{g}lu, and M. Merad, {\it Indian J. Phys.} \textbf{98} (2024) 4093.

\bibitem{Bouguerne2} H. Bouguerne,  B. Hamil, B.C. L\"utf\"uo\u{g}lu, and  M. Merad, {\it Nucl. Phys. B} \textbf{1007} (2024) 116684.

\bibitem{Benchikha} A. Benchikha, B. Hamil, B.C. L\"utf\"uo\u{g}lu, and  B. Khantoul, {\it Int. J. Theor. Phys.} \textbf{63} (2024) 248.

\bibitem{Lut} B.C. L\"utf\"uo\u{g}lu, A. Benchikha, B. Hamil, and B. Khantoul, {\it Mod. Phys. Lett. A} \textbf{40} (2025) 2550009.

\bibitem{Hocine} A. Hocine, F. Merabtine, B. Hamil, B.C. L\"utf\"uo\u{g}lu, and  M. Benarous, {\it  Indian J. Phys.} \textbf{ 99} (2025) 775.

\bibitem{Rou} N. Rouabhia, M. Merad, and B. Hamil, {\it  EPL} \textbf{143} (2023) 52003.

\bibitem{Hamil2} B. Hamil, B.C. L\"utf\"uo\u{g}lu, {\it  Physica A} \textbf{ 623} (2023) 128841.

\bibitem{Hamil3}B. Hamil, B.C. L\"utf\"uo\u{g}lu, and M. Merad, {\it Mod. Phys. Lett. A}, {\it Dunkl-Klein-Gordon equation in higher dimensions},
https://doi.org/10.1142/S0217732325500944 (to be published).

\bibitem{shihai} G.H. Sun, K.D. Launey, T. Dytrych, S.H. Dong, and J.P. Draayer, {\it Adv. Math. Phys.} \textbf{1} (2014) 987376.

\bibitem{shihai2} X.D. Song, G.H. Sun, and S.H. Dong, {\it Phys. Lett. A} \textbf{379} (2015) 1402.

\bibitem{baltenkov} A. Baltenkov, and A. Msezane, {\it Eur. Phys. J. D} \textbf{70} (2016) 81.

\bibitem{chungx} W.S. Chung, and H. Hassanabadi, {\it Mod. Phys. Lett. A} \textbf{34}  (2019) 1950190.

\bibitem{chung2} W.S. Chung, A. Schulze-Halberg, and H. Hassanabadi, {\it Eur. Phys. J. Plus} \textbf{138} (2023) 66.

\bibitem{lebedev}N.N. Lebedev, {\it Special Functions and their applications}, Prentice-Hall, New Jersey, 1965.

\bibitem{nikiforov}A.F. Nikiforov, {\it Special Functions of Mathematical Physics}, Birkh\"auser, Basel, 1988.

\bibitem{jahnke} E. Jahnke, and F. Emde, {\it Tables of Functions with Formulae and Curves}, Dover Publications, Inc., New York, 1945.

\bibitem{abramowitz} M. Abramowitz, and I.A. Stegun, {\it Handbook of Mathematical Functions with Formulas, Graphs, and Mathematical Tables}, Dover Publications, Inc., New York, 1965.

\bibitem{jackson} J.D. Jackson, {\it Classical Electrodynamics}, John Wiley \& Sons, New York, 1975.

\bibitem{SCH} A. Schulze-Halberg, {\it  Phys. Scr.} \textbf{97} (2022) 085213.

\bibitem{physicaA} R.D. Mota, D. Ojeda-Guill\'en, and M.A. Xicot\'encatl, {\it Physica A} \textbf{635} (2024) 129525.

\bibitem{chung1}W.S. Chung and  H. Hassanabadi, {\it Eur. Phys. J. Plus} \textbf{136}, (2021) 239.

\end{thebibliography}
\end{document}